\documentclass[runningheads]{llncs}
\usepackage{graphicx}
\usepackage[dvipsnames,svgnames,x11names]{xcolor}
\usepackage[utf8]{inputenc}
\usepackage[english]{babel}
\usepackage{amsmath,amssymb}
\usepackage{graphicx}
\usepackage{algorithmic}
\usepackage{algorithm}
\usepackage{setspace}
\usepackage{latexsym}
\usepackage[dvipsnames]{xcolor}
\usepackage{rotating}
\usepackage{pifont}
\usepackage{url}
\usepackage[colorlinks=true,linkcolor=red,          
    citecolor=blue,        
    filecolor=magenta,      
    urlcolor=cyan]{hyperref}
\usepackage{multirow}
\usepackage{appendix}

\usepackage{multicol}
\usepackage{bigints}
\usepackage{booktabs}

\usepackage[markup=underlined]{changes}
\usepackage{todonotes}
\setcommentmarkup{\todo[color={authorcolor!20},size=\scriptsize]{#3: #1}}

\definechangesauthor[color=NavyBlue]{ahmad}

\newcommand{\cmark}{\ding{51}}%
\newcommand{\xmark}{\ding{55}}%
\newcommand{\E}{{\mathbb{E}}} 
\newcommand{\me}{\mathrm{ME}}
\newcommand{\gc}{\mathrm{GC}}

\newcommand{\Ahmad}[1]{\textcolor{blue}{Ahmad: #1}}
\newcommand{\ms}[1]{\textcolor{ForestGreen}{MS: #1}}
\newcommand{\mm}[1]{\textcolor{purple}{#1}}

\newcommand{\lk}[1]{\textcolor{green}{#1}}

\begin{document}
\title{ Distribution Estimation for Probabilistic Loops\thanks{This research was partially supported by the WWTF  grant ProbInG ICT19-018, the ERC consolidator grant ARTIST 101002685,  the FWF research
projects LogiCS W1255-N23 and P 30690-N35, and the TU Wien SecInt doctoral program.}}
%
%
\author{Ahmad Karimi\inst{1}
\and {Marcel Moosbrugger} \inst{2}
\and {Miroslav Stankovi{\v c}} \inst{2}
\and \\{Laura Kov\'acs} \inst{2}
\and {Ezio Bartocci} \inst{2}
\and {Efstathia Bura} \inst{1}
}
%
\authorrunning{Karimi et al.}
%
\institute{Applied Statistics, Faculty of Mathematics and Geoinformation, TU Wien\\
\and Faculty of Informatics, TU Wien\\
\vspace{5mm}
\email{ahmad.karimi@tuwien.ac.at}
}

\maketitle   

%
\begin{abstract}

We present an algorithmic approach to estimate the value distributions of random variables of probabilistic loops whose statistical moments are (partially) known. 
Based on these moments, we apply two  statistical methods, Maximum Entropy and Gram-Charlier series, to estimate the distributions of the loop's random variables.
We measure the accuracy of our distribution estimation by comparing the resulting distributions using exact and estimated moments of the probabilistic loop, and  performing statistical tests. 
We evaluate our method on several probabilistic loops with polynomial updates over random variables drawing from common probability distributions, including examples  implementing financial and biological models. 
For this, 
we leverage symbolic approaches to compute  exact higher-order moments of loops
as well as use sampling-based techniques to estimate moments from loop executions. 
Our experimental results  provide practical evidence of the  accuracy of our method for estimating distributions of probabilistic loop outputs. 

\end{abstract}



\section{Introduction}\label{section:Introduction}


Probabilistic programs (PPs) are programs 
with primitives to draw from probability distributions.
As such, PPs do not produce a single output but rather a probability distribution over outputs. In consequence, PPs provide a powerful framework to model system behavior  involving 
uncertainties. 

Many machine and statistical learning techniques leverage PPs for representing and updating data-driven AI systems~\cite{Ghahramani2015}. 
Quantifying and modelling  distributions arising from PPs, and thus formally capturing PP behavior, is challenging. In this work, we address this challenge and provide an algorithmic approach to effectively estimate the probability distributions of  random variables generated by PPs, focusing on PPs with unbounded loops.

While sampling-based techniques are standard statistical approaches to approximate probability distributions in PPs, see e.g.~\cite{Hastings1970}, they cannot be applied to infinite-state PPs with potentially unbounded loops. More recently, static program analysis was combined with statistical techniques to infer  higher-order statistical moments of random program variables in PPs with restricted  loops and polynomial updates~\cite{bartocci2019automatic,Moosbruggeretal2022}. Our work complements these techniques with an algorithmic approach to compute the  distributions of random variables in PPs with unbounded loops for which (some)  higher-order moments are known (see Algorithm~\ref{fig:algo}). 
Moreover, we assess the quality of our estimation via formal statistical tests.

%

Our method can be applied to any PP for which some moments are known. We provide an algorithmic solution to the so-called finite-moment problem~\cite{gavriliadis2009moment,john2007techniques}, as follows: {\it using a finite number of statistical moments of a PP random variable,  compute  the probability density function (pdf) of the variable to  capture the probability of the random variable falling within a particular range of values}. 
In full generality, the finite-moment problem is ill-posed and  there is no single best technique available to solve it~\cite{DeSouzaetal2010}. In our approach, we tackle the finite-moment problem for PPs by focusing on two particular statistical methods, the
Gram-Charlier expansion~\cite{hald2000early} and Maximum Entropy~\cite{biswas2010function,lebaz2016reconstruction},  to estimate the distribution of PP variables. Our approach is further complemented by statistical goodness-of-fit tests for assessing the accuracy of our estimated pdfs, such as the chi-square~\cite{ross2020introduction} and Kolmogorov-Smirnov~\cite{massey1951kolmogorov} tests.



%
%
%
%

\begin{figure}[t]
    \centering
    \begin{minipage}{0.4\textwidth}
    \begin{algorithmic}
        \STATE $a:=0.5$
        \STATE $b:=0.2$
        \STATE $\sigma:=0.2$
        \STATE $w:=0$
        \STATE $r:=2$
        \WHILE{true} 
            \STATE $w:=\text{Normal}(0,1)$
            \STATE $r:=(1-a) r + a b + \sigma w$
        \ENDWHILE
    \end{algorithmic}
    \end{minipage}
    \begin{minipage}{0.58\textwidth}
        \includegraphics[width=6.5cm]{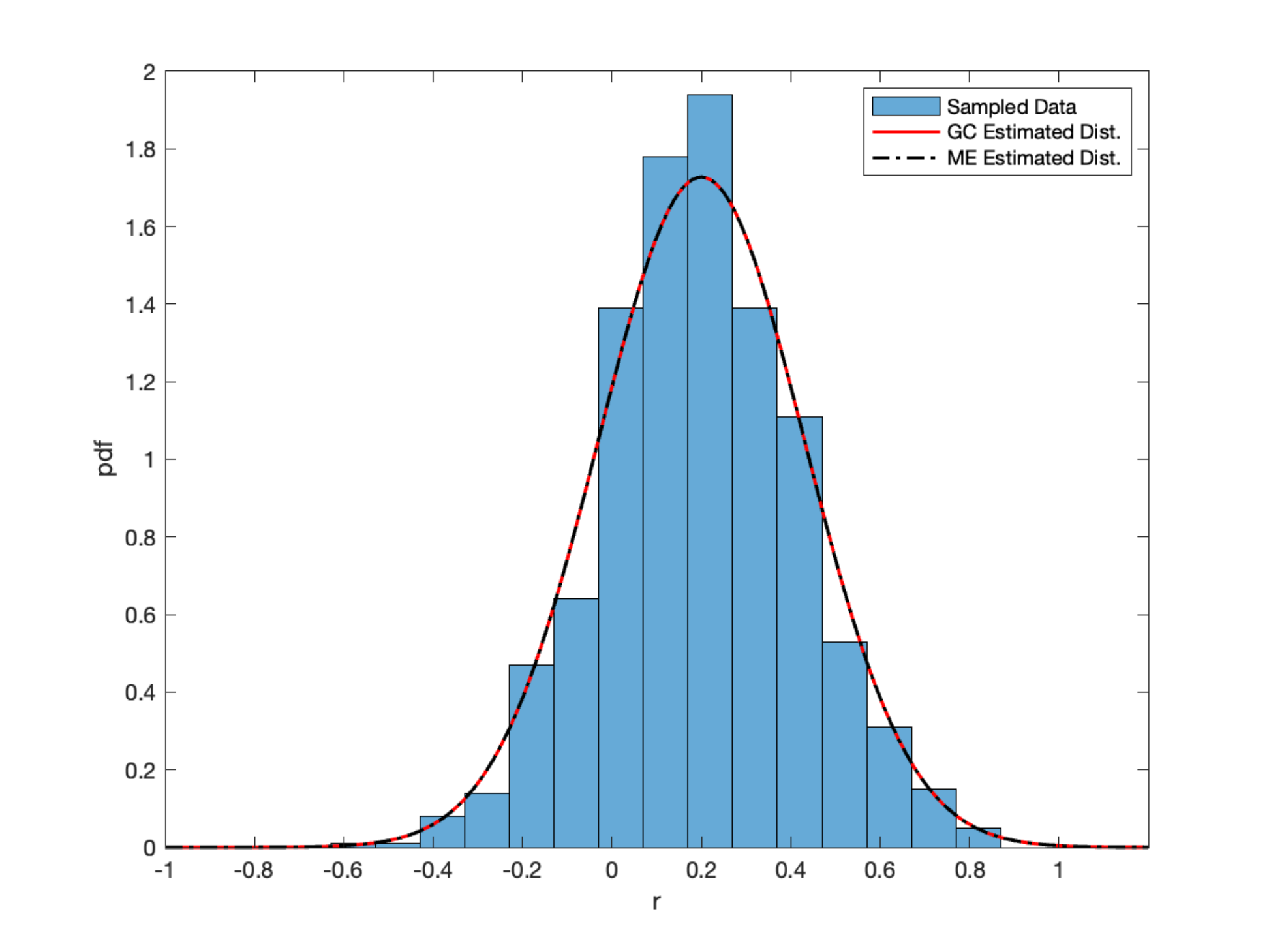}
    \end{minipage}
    \caption{The Vasicek model~\eqref{vasicek} describing the evolution of interest rates, modeled as a PP in the left panel. 
    The estimated distributions  of $r$ are plotted in the right panel. The (normalized) histogram plots the 1000 sampled  $x$-values generated by running the PP 1000 times at iteration $n=100$. The kernel density estimates of the  Maximum Entropy (black dash-dotted line) and 
    Gram-Charlier (red solid line) pdf estimations follow closely the sampled data based histogram, showcasing that both estimates are  effectively estimating the true (sampled) distributions of $r$ (blue histogram). 
 %
 %
    }
    \label{fig:vasicek-intro}
\end{figure}

\paragraph{\bf Motivating Example.} 
The Vasicek model in finance~\cite{vasicek1977equilibrium} describes the evolution of interest rates and serves to motivate our work. This model is defined by the stochastic differential equation, 
\begin{equation}
 dr_{t}=a(b-r_{t})\,dt+\sigma \,dW_{t}   \label{vasicek}
\end{equation}
where $W_t$ is a Wiener process (standard Brownian motion) \cite{morters2010brownian} modeling the continuous inflow of randomness into the system, $\sigma$ is the standard deviation representing the amplitude of the randomness inflow, $b$ is the long term mean level around which paths evolve  in the long term, $a$ is the speed of reversion specifying the velocity at which such trajectories will regroup around $b$ in time.
We encode~\eqref{vasicek} as a PP in Fig.~\ref{fig:vasicek-intro}, using program constants $a,b,\sigma$, as described above and variables $r,w$ to respectively capture the randomness ($r_t$) and Wiener process ($W_t$) of the Vasicek model. The PP of Fig.~\ref{fig:vasicek-intro} has polynomial loop  updates over random variables $r,w$ drawn from a  normal  distribution with 
zero mean and unit variance. This PP satisfies the programming model of~\cite{bartocci2019automatic,Moosbruggeretal2022}, and as such higher-order moments of this PP 
can be computed using~\cite{bartocci2019automatic,Moosbruggeretal2022}. Based on the first- and second-order 
moments of $r$ in the Vasicek model PP, 
we estimate the distribution of the random variable $r$ using 
 Maximum Entropy  and Gram-Charlier expansion (see Section~\ref{ssec:reconstructing-distribution}). Data generated by executing the PP repeatedly, at loop iteration 100, are summarized in the histogram in the right panel of Fig.~\ref{fig:vasicek-intro}, giving a rough estimate  of the distribution of $r$. The black and red lines are kernel density estimates~\cite{Silverman1998} of the pdf of $r$ using 
Maximum Entropy and Gram-Charlier expansion, respectively. Both pdf estimates  closely track the histogram, a proxy for the true distribution of $r$. 
The close match between our estimated pdfs and the true distribution of $r$ is also supported by the chi-square and Kolmogorov-Smirnov test results (see Section~\ref{ssec:methods-of-evaluation}).
%


\paragraph{\bf Related Works.} Kernel density estimation is combined with a constrained minimization problem for an (Hausdorff) instance of the  finite-moment problem in~\cite{athanassoulis2002truncated}. This approach is further extended in~\cite{gavriliadis2009moment}. 
However, this nonparametric estimation method requires constant tuning of the bandwidth parameters, which increase with the number of moments, and is seen to often perform poorly~\cite{lebaz2016reconstruction}.
Along similar lines, the approach of~\cite{john2007techniques} employs spline-based non-parametric density estimation using piecewise polynomial functions to recover  target distributions. This method is  improved in~\cite{DeSouzaetal2010} to an adaptive algorithm dealing with non-equidistant grids. 
This method, though, assumes a smooth transition at the boundaries of sub-intervals that needs to be checked repeatedly during the reconstruction process \cite{lebaz2016reconstruction}. 
Our work complements these approaches by a tailored algorithmic method to effectively  estimate distributions arising from PPs that  uses higher-order moments of these PPs in combination with Maximum Entropy  and Gram-Charlier expansion.
To the best of our knowledge, 
these statistical methods  have not yet been used and evaluated in the setting of probabilistic loops.


Our work is related to emerging efforts in estimating probabilistic distributions/densities in PP, as addressed by the works~\cite{gehr2016psi,gehr2020lpsi,HuangDM21,HoltzenBM20}.
The provided automation, supported by the tools ($\lambda$)PSI \cite{gehr2016psi,gehr2020lpsi}, AQUA~\cite{HuangDM21}, and DICE~\cite{HoltzenBM20}, yield  symbolic frameworks to  compute exact (posterior) densities for PPs with {bounded loops}. Also focusing on bounded loops,  sampling-based techniques to approximate the distributions resulting from PPs are exploited in~\cite{CaretteS16,NarayananCRSZ16,Carpenter2017StanAP,BhatBGR17}.
In contrast, our approach is not restricted to bounded loops but focuses on density estimation for PPs with unbounded loops.


\paragraph{\bf Our Contributions.}
The  main contribution of this paper is the development of  an {\it algorithmic  approach to effectively estimate distributions arising from PPs with unbounded loops (Algorithm~\ref{fig:algo})}. We employ two formal techniques from mathematical statistics, Maximum Entropy (ME) and Gram-Charlier (GC) expansion, to estimate distributions of PP variables (Section~\ref{ssec:reconstructing-distribution}).  We use symbolic approaches to compute exact moments of loop variables as symbolic expressions parameterized by the loop counter.
Further, we apply statistical tests to assess the adequacy of the estimated distributions of PP variables (Section~\ref{ssec:methods-of-evaluation}). We evaluate our approach on a number of benchmarks and demonstrate the accuracy of our proposed estimation approach (Section~\ref{section:ImplEvaluation}).


\paragraph{\bf Paper Outline.} The rest of the paper is organized as follows.  Section~\ref{section:Preliminaries} introduces the necessary prerequisites from probability theory, statistics, and probabilistic programs.
In Section~\ref{section:Methodology}, we introduce the (adjusted) methods we use to estimate the distributions of program variables from their moments.
We report on our practical findings in Section~\ref{section:ImplEvaluation}, and conclude the paper in Section~\ref{section:Conclusion}. 


\section{Preliminaries}\label{section:Preliminaries}
This section reviews relevant terminology from statistics and probabilistic programs; for further details we refer to~\cite{durrett2019probability,Moosbruggeretal2022}. 
Throughout the paper, $\mathbb{N}$ denotes the set of natural numbers. To facilitate readability,  we sometimes write $\exp[t]$ to denote the exponential function $e^{t}$, where $t$ is an arbitrary expression/argument.

\subsection{PPs and Moments of Random Variables}
{ The result of a PP is not a single output but rather multiple values  with different probabilities according to the distribution of the random variables the PP encodes. }
Different execution paths in a
PP are typically selected by draws from commonly used distributions that are fully characterized by their moments, such as the uniform or  normal. 
Yet, since  output values of PPs
are  results of multiple operations,  distributions arising from the random variables the PPs encode are most often not common.

\paragraph{\bf Moments of Probabilistic Loops.}


Moments of program variables of probabilistic loops can be approximated by sampling; that is, by executing the loops for an arbitrary but fixed number of iterations (see, e.g.~\cite{Hastings1970,younes2006statistical}).
Alternatively, 
for restricted classes of probabilistic loops with polynomial updates, symbolic methods from algorithmic combinatorics can be used to compute  the exact 
(higher-order) moments of random program variables $x$ by expressing these moments as closed-form expressions over loop iterations and some initial values~\cite{bartocci2019automatic,Moosbruggeretal2022}. 
That is, ~\cite{bartocci2019automatic,Moosbruggeretal2022} derive the expected value 
of the $k$th moment of variable $x$ at loop iteration $n$, denoted as $\mathbb{E}(x^k(n))$, 
in closed form, where $x^k(n)$ specifies the value of $x^k$ at loop iteration $n$, and $k,n\in\mathbb{N}$. 

\subsection{From Moments to Distributions}\label{section:FromMomentToDistribution}

Given a finite set of moments of a random variable $x$,  its distribution can be estimated using various statistical approaches. We focus on those we use herein.

\paragraph{\bf Maximum Entropy (ME).}\label{section:MaximumEntropyMethod}
The Maximum Entropy (ME) distribution estimation  method is based on the maximization of constraints describing the Shannon information entropy~\cite{biswas2010function,buchen1996maximum,lebaz2016reconstruction}.  
%
Specifically, in order to estimate the unknown distribution $f$ of a PP variable $x$ in our setting, we  maximize the {Shannon entropy $H$ of $f$, defined by} 
\begin{align}\label{ShanonEntropy}
  H[f]=-\int_{l}^{u} f(x) \ln(f(x))dx,  
\end{align}
subject to its given moments $\E(x^i)=\int x^if(x)dx$ \cite{bernardo2009bayesian,Kendall2004}.
The ME approximation $f_{\me}(x)$ of the target 
probability density function (pdf) takes the form 
\begin{equation}\label{eq:pdf:ME}
f_{\me}(x)=\exp\left[-\sum_{j=0}^N\xi_jx^j\right],
\end{equation}
where the Lagrange's multipliers $\xi_j$, with $j=0,1,\dots,m$, 
can be obtained from  the first $m\in\mathbb{N}$ 
moments, $\{\E(x),\E(x^2),\dots,\E(x^{m})\}$. To this end, the following system of  $m+1$ 
nonlinear equations is solved,
\begin{align}\label{eq:nonlinearSystemME}
\bigintss_l^u x^i\exp\left[-\sum_{j=0}^{m}\xi_jx^j\right]dx=\E(x^i).   
\end{align}

\subsubsection{Gram-Charlier Expansion}\label{section:GramCharlierExpansion}
The Gram–Charlier  series   approximates the pdf $f$ of a PP variable $x$ in terms of its cumulants {and using a known distribution $\psi$}~\cite{hald2000early,brenn2017revisit}. We let $\psi$ to be a normal/Gaussian  distribution (see Section~\ref{ssec:reconstructing-distribution}).  As an alternative to moments,  cumulants  of a distribution are defined using the {\it{cumulant-generating function}}, which equals the natural logarithm of the characteristic function, $K(t)=\ln\left(\E(e^{itx})\right)$. In what follows, we denote by $\kappa_m$ the $m$th cumulant of $f$, the unknown target distribution  to be approximated.  The relationship between  moments and cumulants can be obtained by extracting coefficients from the expansion. 
To be precise, we can express the $m$th cumulant $\kappa_m$ in terms of the first $m$ moments~\cite{broca2004cumulant} as 
\begin{equation}\label{eq:CumulantMomentRelation}
\footnotesize{
    \kappa_m=(-1)^{m+1}\det
    \begin{pmatrix}
    \E(x)& 1 & 0 & 0 & 0 & \dots & 0\\
    \E(x^2)& \E(x) & 1 & 0 & 0 & \dots & 0\\
    \E(x^3)& \E(x^2) & \binom{2}{1}\E(x) & 1 & 0 & \dots & 0\\
    \E(x^4)& \E(x^3) & \binom{3}{1}\E(x^2) & \binom{3}{2}\E(x) & 1 & \dots & 0\\
    \vdots& \vdots & \vdots & \vdots & \vdots  & \ddots & \vdots\\
    \E(x^{m-1})& \E(x^{m-2}) & \dots & \dots & \dots & \ddots & 1\\
    \E(x^m)& \E(x^{m-1}) & \dots & \dots & \dots & \dots & \binom{m-1}{m-2}\E(x) \\
    \end{pmatrix},
    }
\end{equation}
where $\det(\cdot)$ stands for determinant.
The first cumulant  $\kappa_1$  of the random variable $x$ is the mean $\mu=\E(x)$; the second cumulant $\kappa_2$ of $f$ is the variance $\sigma^2$, and the third cumulant $\kappa_3$ is the same as the third central moment\footnote{The $i$th central moment of $x$ is defined as $\E\left((x-\E(x))^i\right).$}. 
Higher-order cumulants of $f$, however, are in general not equal to higher moments.  
Using the cumulants $\kappa_m$ computed from exact moments of $x$ together with  the cumulants 
of the known distribution $\psi$, 
the pdf $f$ of a random variable $x$  can approximated by the Gram-Charlier (type-)A 
expansion $f_{\gc}(x)$, as in~\cite{KendallStuart1977} and given by 
\begin{align}\label{pdfGCGeneralC}	
f_{\gc}(x) = 
 {\psi}(x) \sum_{m=0}^{\infty}\frac{1}{m!\sigma^m}B_m(0,0,\kappa_3,\dots,\kappa_m) He_m \left(\frac{x-\mu}{\sigma}\right),
\end{align} 
where $\psi$ is the normal pdf  with mean $\mu=\kappa_1$ and variance $\sigma^2=\kappa_2$, and $B_m$ and $He_m$ are respectively the Bell and  Hermite polynomials~\cite{brenn2017revisit}. Derivation details of \eqref{pdfGCGeneralC} can be found in \cite{wallace1958asymptotic}.

\section{Effective Estimation of Distributions for Probabilistic Loops}\label{section:Methodology}
%

We present our estimation approach for the pdf of a random variable $x$ 
in a probabilistic loop $\mathcal{P}$, provided the first $M$ statistical moments of $x$ are known. We use 
Maximum entropy (ME) and Gram-Charlier A (GC)  expansion (lines~{1--4} of Algorithm~\ref{fig:algo}). 
We  assess the accuracy of our pdf estimates by conducting statistical tests over our distribution estimates (lines~5--14 of  Algorithm~\ref{fig:algo}). 
Our approach is summarized in Algorithm~\ref{fig:algo} and detailed next.

\begin{algorithm}[t!]
\begin{algorithmic}[1]
    \REQUIRE Probabilistic loop $\mathcal{P}$ with program variable(s) $x$;\\ 
    \qquad set $M$ of exact  moments of~$x$ for loop iteration $n$ of $\mathcal{P}$\\
    \ENSURE Estimated distributions  $f_{\me}$ and $f_{\gc}$ of $x$,  with respective {accuracy} $\mathcal{A}_{\me}$ and $\mathcal{A}_{\gc}$
    \\
    \hspace*{-1.7em}{\bf Parameters:} Loop iteration $n\in\mathbb{N}$; number of executions $e\in \mathbb{N}$  of $\mathcal{P}$\\
    \vspace{0.5em} \textbf{Initialization:}\\
    \STATE Choose a subset of exact moments $S_{\text{M}}=\{\E(x),\E(x^2),\dots,\E(x^{|S_{\text{M}}|})\} \subset M$ 
    \STATE Collect ${Sample_{Data}}$ by sampling $e$ many $\mathcal{P}$ variable values at the $n$th loop iteration \\
    \vspace{0.5em}
    \textbf{Distribution Estimation:}\\
    \STATE Compute $f_{\me}$ using $S_{\text{M}}$ and ME as in~\eqref{eq:pdf:ME}\\
    \STATE Compute $f_{\gc}$ using  $S_{\text{M}}$ and GC as in~\eqref{pdfGC}\\

    \vspace{0.5em}
    \textbf{$\boldsymbol\chi^2$ Test:}\\
    \STATE Split ${Sample_{Data}}$ into bins and compute observed frequencies $O_i$\\
    \STATE Calculate expected frequencies $E_{\me, i}$ and $E_{\gc, i}$ from $f_{\me}$ and $f_{\gc}$, as in~\eqref{eq:ChiSquare:EFreqs}\\
    \STATE Compute $\chi^2$ test statistics for $f_{\me}$ and $f_{\gc}$ as in~\eqref{eq:ChiSquare:Tests} and compare them to the critical value $CV_{\chi^2}$\\
    \vspace{0.5em} \textbf{K-S Test:}\\
    \STATE Calculate the empirical cdf $F_{Sample}$ of $Sample_{Data}$ and cdfs $F_{\me}$, $F_{\gc}$ using $f_{\me}$ and $f_{\gc}$\\
    \STATE Compute K-S test statistics $D_{\me}^\ast$, $D_{\gc}^\ast$ of  $f_{\me}$ and $f_{\gc}$ as in~\eqref{eq:KSTest} and compare to the critical value $CV_{\texttt{K-S}}$\\
    
    
    \vspace{0.5em} \textbf{Pdf Accuracy Evaluation:}\\
    \STATE If $\chi_{\me}^2<CV_{\chi^2}$ or $D_{\me}^\ast<CV_{\texttt{K-S}}$ then $\mathcal{A}_{\me}\leftarrow NOT\; REJECTED$ \\
    \qquad\qquad\qquad\qquad\qquad\qquad\qquad\quad else $\mathcal{A}_{\me}\leftarrow REJECTED$
    \STATE If $\chi_{\gc}^2<CV_{\chi^2}$ or $D_{\gc}^\ast<CV_{\texttt{K-S}}$ then $\mathcal{A}_{\gc}\leftarrow NOT\; REJECTED$ \\
    \qquad\qquad\qquad\qquad\qquad\qquad\qquad\quad else $\mathcal{A_{\gc}}\leftarrow REJECTED$
\end{algorithmic}
\caption{Effective Distribution Estimation}
\label{fig:algo}
\end{algorithm}

\subsection{Distribution Estimation}\label{ssec:reconstructing-distribution}

Given a variable $x$ of a probabilistic loop $\mathcal{P}$, we use a subset $S_{\text{M}}\subset M$  of its exact moments for estimating the  distribution of $x$  through ME and GC expansion. 
For this, 
we  adjust ME and GC expansion  
to compute the  estimated distributions $f_{\gc}$ and $f_{\me}$ of $f$, the pdf of $x$, respectively. 
For doing so, we use the set $S_{\text{M}}=\{\E(x),\E(x^2),\dots,\E(x^{|S_{\text{M}}|})\}$ of exact moments of $x$. 
For ME, we derive the Lagrange multipliers $\xi_j$, for $j=0, 1, \ldots, |S_{\text{M}}|$ in the ME approximation~\eqref{eq:pdf:ME}, where $|S_{\text{M}}|$ denotes the cardinality of $S_M$.  %
%
%
For GC expansion, we truncate the GC expansion~\eqref{pdfGCGeneralC} based on $|S_M|$ moments, and we use the $f_{\gc}(x)$  to estimate the pdf of $x$ as
\begin{align}\label{pdfGC}	f_{\gc}(x) = 
{\psi}(x) \sum_{m=0}^{|S_{\text{M}}|}\frac{1}{m!\sigma^m}B_m(0,0,\kappa_3,\dots,\kappa_m) He_m \left(\frac{x-\mu}{\sigma}\right).
\end{align} 
The cumulants $\kappa_m$ of the pdf of $x$ in \eqref{pdfGC} are computed from the  moments in  $S_{\text{M}}$. 

The $f_{\gc}(x)$ estimate of the pdf of a PP variable~$x$ can be computed 
even when the moments of~$x$ are parametric; i.e., their closed-form functional representations depend on the loop iteration~$n$ and/or other symbolic values. This is especially useful in the analysis of (probabilistic) loops, as it allows us to encompass {\it all} loop iterations in a single symbolic estimate using GC expansion. 

\begin{example}\label{exactmomentsVasicek} We use the PP in Fig.~\ref{fig:vasicek-intro} to illustrate our approach.
The set $M$ in Algorithm~\ref{fig:algo} contains the exact first two moments of $r$ for an arbitrary loop iteration $n$. These two moments can be expressed as functions of loop iteration~$n$ of the PP (see Sec.~\ref{section:ImplEvaluation}), 
    \begin{equation}\label{eq:Vasicek:R:2Moments}
        \begin{split}
            &\E(r(n)) = 2^{-n}(2^n + 9)/5,\\
            &\E(r^2(n)) = 7/75 + (18)2^{-n}/25 + (239)2^{-2n}/75.
        \end{split}
    \end{equation}
   As a result, we set $S_{\text{M}}=M=\{\E(r(n)),\E(r^2(n))\}$ for loop iteration $n$ (line~1 of Algorithm~\ref{fig:algo}). These functions yield exact moments of  $r$ for concrete values of the loop iteration $n$, by only instantiating the above expressions with the respective values of~$n$. For example, at 
    loop iteration $n=100$, the two exact moments of $r$ are $\E(r(100)) = 0.20 $ and $\E(r^2(100)) = 0.093\bar{3} $.


\paragraph{ME estimation of the pdf of $r$.} 
Given~\eqref{eq:Vasicek:R:2Moments}, 
we solve the  system~\eqref{eq:nonlinearSystemME}
of nonlinear equations to obtain the respective  Lagrange multipliers $\xi_j$, $j = 0,1,2$. 
For solving~\eqref{eq:nonlinearSystemME}, 
we assume that the support of  the target pdf of $r$ is a subset of $[l,u]$, where $l$ and $u$ are known scalars and apply the Levenberg–Marquardt numerical minimization algorithm~\cite{levenberg1944method,marquardt1963algorithm}. 
Once the optimal multipliers for~\eqref{eq:nonlinearSystemME} are computed, the ME estimate is $f_{\me}(r)=\exp\left[-(\xi_0+\xi_1r+\xi_2r^2)\right]$. 
Specifically, $f_{ME}(r)= \exp[0.171658 + 3.749999 \,r + 9.374997\, r^2]$ at loop iteration $n=100$. As seen from Fig.~\ref{fig:vasicek-intro}, this estimate is closely approximating the true (sampled) distribution of $r$.



\paragraph{GC estimation of the pdf of $r$.}  Given the first two exact moments of $r$ in \eqref{eq:Vasicek:R:2Moments}, we apply ~\eqref{eq:CumulantMomentRelation} in order to compute the corresponding cumulants.  
Using the normal distribution  with mean $\mu=\kappa_1=\E(r)$ and variance $\sigma^2=\kappa_2=\E(r^2)-\E^2(r)$ in~\eqref{pdfGC}, 
the GC estimate of the pdf  of $r$ is $f_{\gc}(r)=({1}/{\sqrt{2\pi\kappa_2}})\exp\big[-{(r-\kappa_1)^2}/{2\kappa_2}\big]$. 
By expressing  $\kappa_1$ and $\kappa_2$ in terms of exact moments,  $\kappa_1=\E(r)$ and $\kappa_2=\E(r^2)-\E^2(r)$, and using~\eqref{eq:Vasicek:R:2Moments}, 
the GC estimate is further expressed as a function of $n$: 
{\begin{align*}
        f_{\gc}(r(n))=
        &\frac{\eta\; \exp\left[-\frac{\left(r(n) - \frac{2^n + 9}{5 \cdot  2^n}\right)^2}{\frac{36 \cdot 2^{-n}}{25} 
        - \frac{2 \cdot (2^n + 9)^2}{2^{2n} \cdot 25} + \frac{478}{2^{2n} \cdot 75} + \frac{14}{75}}\right]}{\beta\; \left(\frac{18 \cdot2^{-n}}{25}
        - \frac{2^{-2n}(2^n + 9)^2}{25} + \frac{239 \cdot 2^{-2n}}{75} + \frac{7}{75}\right)^{1/2}}~,
    \end{align*}}
\noindent where $\eta=2251799813685248$ and $\beta=5644425081792261$. 
In particular, for loop iteration $n=100$, the obtained  $f_{\gc}(r)=1.7275\exp[-9.3750(r - 0.2)^2]$  closely approximates the true (sampled) distributed of $r$, as evidenced in Fig.~\ref{fig:vasicek-intro}.



\end{example}

\subsection{Assessing Accuracy of Estimated Distributions}\label{ssec:methods-of-evaluation}

To evaluate the accuracy of the estimated pdfs $f_{\me}$ and $f_{\gc}$ of the PP variable~$x$, we would, ideally, compare them to the true underlying pdf of~$x$. This comparison, however, is not possible in general, as the true distribution of $x$ arising from an (unbounded) probabilistic loop is complex and unknown. To overcome this obstacle, we 
execute the  probabilistic loop $e$ times to  
sample the distribution of~$x$ and use the such collected $Sample_{Data}$ as proxy of the underlying probability distribution of $x$ (line~2 of Algorithm~\ref{fig:algo}). For this, we carry out a Monte-Carlo ``experiment''~\cite{Hastings1970},  by executing $\mathcal{P}$ a large number of times ($e$) and collect the value of $x$ at loop iteration $n$ in $Sample_{Data}$\footnote{in our experiments, we use $e=1000$ and $n=100$, see Section~\ref{section:ImplEvaluation}}. 
%
Based on $Sample_{Data}$, we  evaluate the accuracy of our ME and GC expansion estimations using {two} statistical tests, namely the 
chi-square ($\chi^2$)~\cite{ross2020introduction} and the Kolmogorov-Smirnov~\cite{massey1951kolmogorov} goodness-of-fit tests, as described next. 

\paragraph{{\bf Chi-square ($\chi^2$) Goodness-of-Fit Test.}} The chi-square ($\chi^2$) goodness-of-fit test is a common statistical test to detect statistically significant differences between expected and observed frequencies~\cite{Snedecor98},  
by testing if a sample comes from a specific distribution.   We use  the $\chi^2$ goodness-of-fit test to compare $f_{\me}$ and $f_{\gc}$ with the ``true'' (sampled) empirical distribution of $x$ that is based on the sampled data, $Sample_{Data}$ (lines~5--7 of Algorithm~\ref{fig:algo}). 

We partition our $Sample_{Data}$  into~$k\in\mathbb{N}$ non-overlapping intervals $I_k$ (also called  bins), such that $Sample_{Data}=\bigcup_{i=1}^k I_i$.
Let 
$O_i=|I_i|$ denote the number of samples in the $i$th interval $I_i$, with  $i\in\{1,\ldots,k\}$, and let $l_i$
 and $u_i$ respectively denote the lower and upper bounds of $I_i$. As such,  $O_i$ is the \textit{observed} data frequency of $x$ for interval $I_i$.    
We compute the expected frequencies\footnote{alternatively, these frequencies can also be obtained from the 
cumulative distribution function (cdf) of $x$
} in $I_i$, denoted as $E_{\me,i}$ and $E_{\gc,i}$,  from  $f_{\me}$ and $f_{\gc}$  as 
\begin{equation}\label{eq:ChiSquare:EFreqs}
\begin{array}{l}
E_{\me,i}=|Sample_{Data}|*\int_{l_i}^{u_i} f_{\me}(x) dx, \\[.5em]
E_{\gc,i}=|Sample_{Data}|*\int_{l_i}^{u_i} f_{\gc}(x) dx.
\end{array}
\end{equation}
The resulting chi-square goodness-of-fit test statistics are
\begin{equation}\label{eq:ChiSquare:Tests}
\begin{array}{ll}
    \chi_{\me}^2=\sum_{i=1}^k (O_i-E_{\me,i})^2/E_{\me,i},\\[.5em]
        \chi_{\gc}^2=\sum_{i=1}^k (O_i-E_{\gc,i})^2/E_{\gc,i}.\\[.5em]
\end{array}
\end{equation}
If these values exceed the chi-square critical value $CV_{\chi^2}=\chi^2_{1-\alpha,k-1}$, where $\alpha$ is the statistical significance level (parameter) and $k-1$ the degrees of freedom, the hypothesis that the ME or GC estimated distributions, respectively, are the same as the ``true'' distribution of $Sample_{Data}$ is rejected (lines~10--11 of Algorithm~\ref{fig:algo}). 
Otherwise, there is not enough evidence to support the claim that the distributions differ significantly. 


%



\begin{example}\label{example:VasicekModel-Chi2}
We set $k=15$ and $\alpha=0.05$, so that $CV_{\chi^2}=\chi^2_{1-\alpha,k-1}=23.685$.    Figure~\ref{fig:vasicek-histogram-ME-GC-bins}  shows observed  frequencies from $Sample_{Data}$, as well as the expected frequencies from the $f_{\me}$ and $f_{\gc}$ pdfs of Example~\ref{exactmomentsVasicek}, where $Sample_{Data}$ are collected while sampling the PP of Fig.~\ref{fig:vasicek-intro} for $e=1000$ times at the $n=100$th loop iteration.  The test statistic values of \eqref{eq:ChiSquare:Tests} are $\chi_{\me}^2=17.4018$ and $\chi_{\gc}^2=17.4017$. Since both are smaller than $CV_{\chi^2}$, we conclude that both $f_{\me}$ and $f_{\gc}$ are accurate estimates of the pdf of $r$. This result is also supported by  the close agreement of the plotted frequencies in  Fig.~\ref{fig:vasicek-histogram-ME-GC-bins}. 

    
    
\begin{figure}[t!]
	\centering
	\includegraphics[width=8cm]{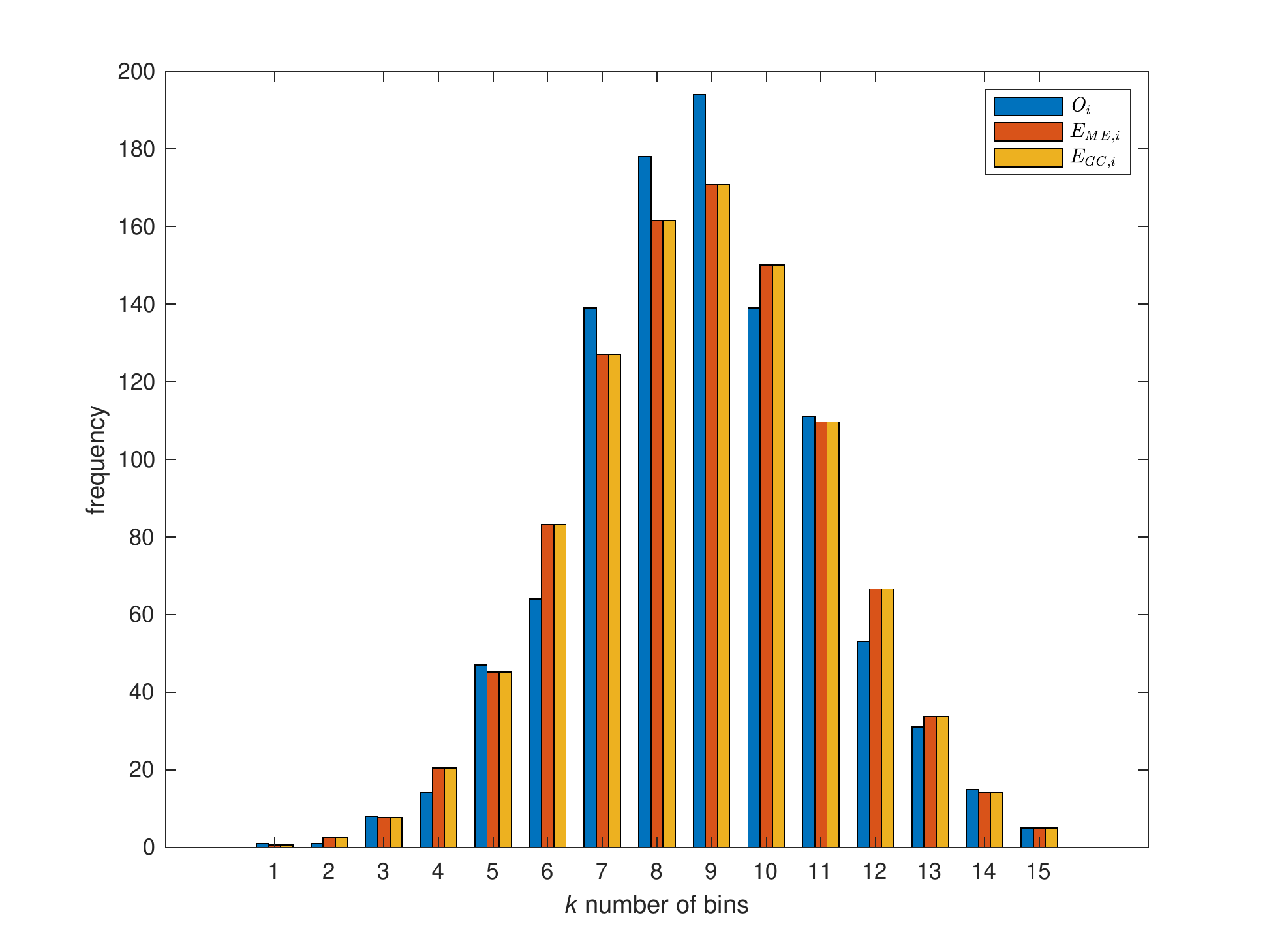}
	\vspace{-5mm}
	\caption{Observed frequencies (blue) of the $Sample_{Data}$ of Figure~\ref{fig:vasicek-intro}, in alignment with the expected frequencies obtained from the $f_{\me}$ (orange) and $f_{\gc}$ (yellow) estimated distributions of the PP of Figure~\ref{fig:vasicek-intro}.}
	\label{fig:vasicek-histogram-ME-GC-bins}
\end{figure}

\end{example}

\paragraph{{\bf Kolmogorov-Smirnov (K-S) Test.}} The Kolmogorov-Smirnov (K-S) test~\cite{massey1951kolmogorov} is a goodness-of-fit test using  cumulative distribution functions (cdfs) to test whether two  distributions differ.   
We compute the cdfs $F_{me}$ and $F_{\gc}$ of the estimated pdfs $f_{\me}$ and $f_{\gc}$, respectively. We also compute the (empirical) cdf~$F_{Sample}$ of $Sample_{Data}$. 
We let 
\begin{equation}\label{eq:KSTest}
    \begin{array}{l}
    D^\ast_{\me}=\max_x(|F_{\me}(x)-F_{Sample}(x)|),\\[.5em]
    D^\ast_{\gc}=\max_x(|F_{\gc}(x)-F_{Sample}(x)|)
\end{array}
\end{equation}
to denote the K-S test statistics $D^\ast_{\me}$ and  $D^\ast_{\gc}$, respectively.
To assess the distance of $F_{\me}$ and $F_{\gc}$ from $F_{Sample}(x)$ with K-S, 
we compare $D^\ast_{\me}$ and $D^\ast_{\gc}$ to the K-S test critical value  $CV_{\text{K-S}}=\sqrt{-({1}/{|Sample_{Data}|})\ln{({\alpha}/{2})}}$, where $\alpha$ is the statistical significance level. The K-S test rejects the claim that the compared distributions are the same if $D^\ast_{\cdot} < CV_{K-S}$ (lines~10--11 of Algorithm~\ref{fig:algo}).


\begin{example}\label{example:VasicekModel-K-S}
We set $\alpha=0.05$, so that $CV_{\text{K-S}}=0.0608$.   We compute the \mbox{K-S} test statistics for the  ME and GC estimated distributions
 of Example~\ref{exactmomentsVasicek}, to obtain
$D^{\ast}_{\me} = 0.03602307$ and $D^{\ast}_{\gc} = 0.03602304$. Since both test statistics are smaller than the critical value, we conclude that both  $f_{\me}$ and $f_{\gc}$ are close to the underlying distribution of $r$. 
 
%
%
\end{example}

\section{Experimental Evaluation}\label{section:ImplEvaluation}
In this section we report on our  experimental results towards estimating distributions of probabilistic loop variables using Algorithm~\ref{fig:algo}. We describe our benchmark set and present our practical findings using these benchmarks. {We also describe additional results on evaluating the precision of higher-order moments of loop variables. }

\begin{table}[t!]
    \centering
     \scalebox{1}{
 \begin{tabular}{c|c|c|l|l|l|l}
\toprule
          {\bf Program}   &  Var & $|S_{\text{M}}|$ & $\chi_{\me}^2$  &$\chi_{\gc}^2$ &  $D_{\me}^\ast$  & $D_{\gc}^\ast$\qquad\\\hline\hline
     StutteringP~\cite{bartocci2019automatic}& $s$ & 2 &18.6432 \cmark& 16.7943 \cmark& 0.0181 \cmark & 0.0213 \cmark \\
     Square~\cite{bartocci2019automatic} & $y$ & 2 &36.9009 \xmark& 39.1309 \xmark &  0.0586 \cmark & 0.0566 \cmark\\
     Binomial~\cite{Feng2017} & $x$ & 2 &27.4661 \xmark& 27.4574 \xmark&  0.0598 \cmark &  0.0597 \cmark \\
   Random Walk 1D~\cite{Kura19} & $x$&  2 &18.6709 \cmark \hspace{2mm}& 18.7068 \cmark \hspace{2mm}&   0.0264 \cmark \hspace{2mm}&  0.0263 \cmark \hspace{2mm}\\
    Uniform(0,1) & $u$ & 6 &16.3485 \cmark& 105.276 \xmark&  0.0214 \cmark & 0.0658 \xmark \\
    Vasicek Model & $r$&  2 &17.4018 \cmark & 17.4017 \cmark & 0.0360 \cmark & 0.0360 \cmark\\
     PDP Model &  $x$ & 3 & 64.4182 \cmark & 65.9500 \cmark &  0.0403 \cmark & 0.0393 \cmark \\\bottomrule
\end{tabular}
}
\vspace{2mm}
 \caption{Accuracy of the ME and GC expansion pdf estimates for  benchmarks PPs, assessed with  
 the chi-square and K-S statistical tests. \label{table:DKL-Chi2-K-SforBenchmarks}}
\end{table}

\begin{figure}[t!]
    \centering
    \begin{minipage}{0.41\textwidth}
\begin{algorithmic}
    \STATE $k_1:=4$, $k_2:=40$, $y:=0$
    \STATE $x:=0$, $a:=0.2$, $b:=4$, $s:=0$
    \STATE  $h:=0.6$, $f:=0.1$, $\rho:=0.5$
    \WHILE{$true$}
        \IF{$s=0$}
            \STATE $s = 1 \;[\;f\;]\; 0$
        \ELSE
            \STATE $s = 0 \;[\;h\;]\; 1$
        \ENDIF
        \STATE $k:= k_2 * s + k_1 * (1-s)$
        \STATE $y:=(1-\rho)y+k$
        \STATE $x:=(1-a)x+by$
    \ENDWHILE
\end{algorithmic}
\end{minipage}
\begin{minipage}{0.58\textwidth}
\includegraphics[width=\textwidth]{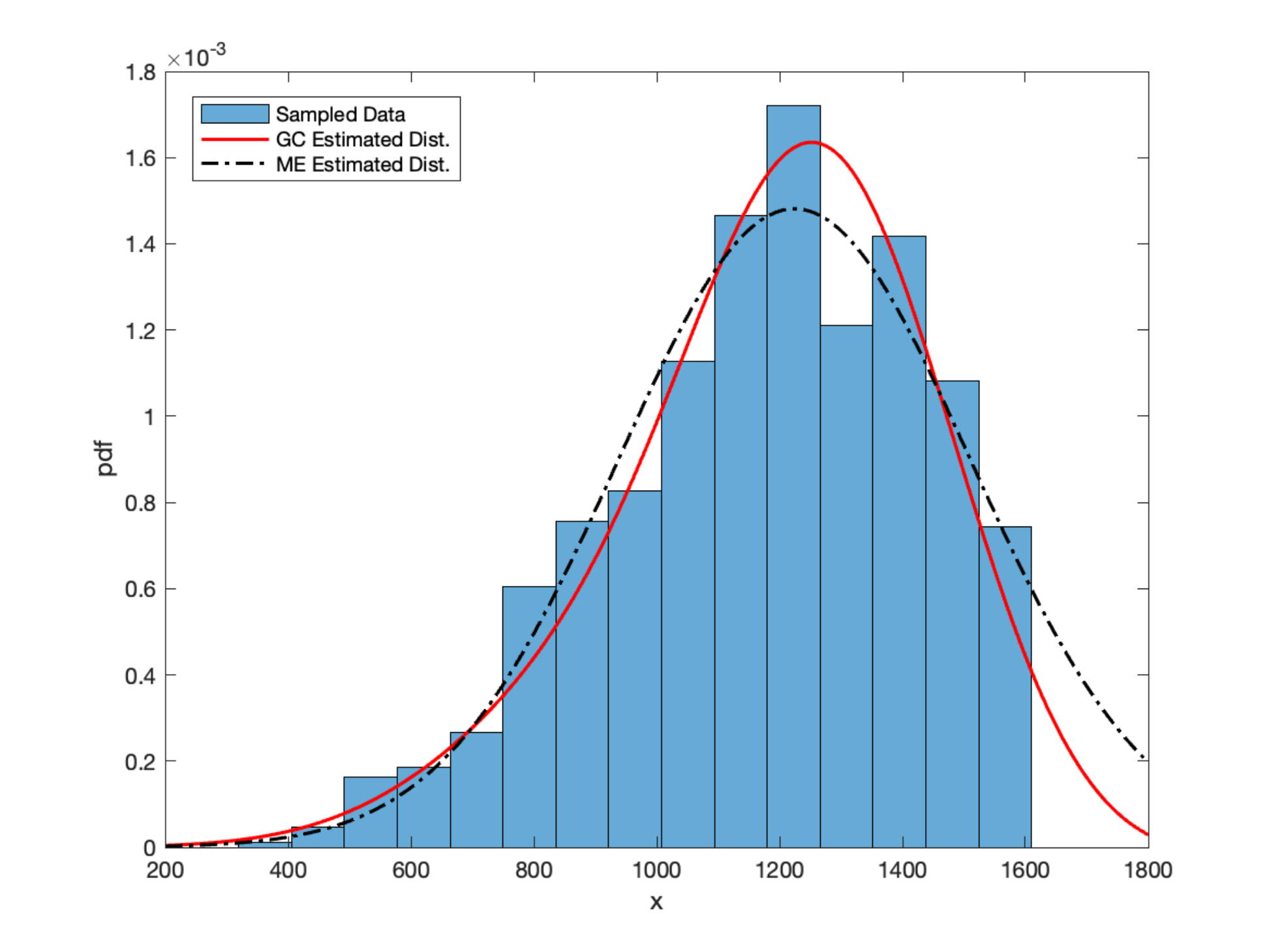}
\end{minipage}
\caption{A PDP model for gene circuits, modeled as a PP in the left,  representing a 
gene  controlled by a two-valued, probabilistically updated variable $s\in\{0,1\}$. 
We respectively denote with $k_0$ and $k_1$ the gene transcription  rates  in states $s=0$ and $s=1$, and let $b$ the translation rate of protein $x$ production in the gene. Further, $\rho$ controls the mRNA level $y$ and $a$ denotes the protein degradation rate.
The (normalized) histogram  of sampled data of the resulting PP, with execution time/sample size $e=1000$ and loop iteration $n=100$ is given on the right, together with the  ME and GC expansion  estimates of the pdf of $x$ using  the first 3 moments of $x$.} 
\label{fig:PDPmodelPPandMoments}
\end{figure}

\paragraph{{\bf Benchmarks.}}
For evaluating our approach in Algorithm~\ref{fig:algo}, we use four  challenging examples of unbounded probabilistic loops from state-of-the-art approaches on quantitative analysis of PPs~\cite{Feng2017,Kura19,bartocci2019automatic}; these benchmarks are the first four entries of the first column of Table~\ref{table:DKL-Chi2-K-SforBenchmarks}. 
In addition, we also crafted three new examples, listed in the last three  entries of the first column of Table~\ref{table:DKL-Chi2-K-SforBenchmarks}, as follows: (i) line~5 of Table~\ref{table:DKL-Chi2-K-SforBenchmarks} specifies a PP loop approximating a uniform distribution; (ii)  line~6 of Table~\ref{table:DKL-Chi2-K-SforBenchmarks} refers to the Vasicek model of Fig.~\ref{fig:vasicek-intro}; and (iii) 
the last line of Table~\ref{table:DKL-Chi2-K-SforBenchmarks} lists an example encoding a piece-wise deterministic process (PDP) modeling gene circuits based on~\cite{innocentini2018time}. In particular, the PDP model we consider 
can be used to estimate the distribution of protein $x$ and the mRNA levels $y$ in a gene; our PP encoding of this PDP model is given in Fig.~\ref{fig:PDPmodelPPandMoments}.

\begin{singlespace}
\end{singlespace}


\paragraph{{\bf Experimental Setup.}}
All our seven examples in Table~\ref{table:DKL-Chi2-K-SforBenchmarks} implement polynomial loop updates and fall in the class of probabilistic loops supported by~\cite{moosbrugger2021automated}. As such, for each example of Table~\ref{table:DKL-Chi2-K-SforBenchmarks}, exact higher-order moments of random loop variables can be computed using the algorithmic approach of~\cite{moosbrugger2021automated}. In our work, we use the Polar tool of~\cite{moosbrugger2021automated} to derive a {\it finite} set $M$ of exact higher-order moments for each PP of Table~\ref{table:DKL-Chi2-K-SforBenchmarks}, and we set $S_M = M$ to be further used in Algorithm~\ref{fig:algo}.
Further, we 
%
generate our sampled data ($Sample_{Data}$)  by executing each PP $e=1000$ times and  for loop iteration $n=100$. 
For assessing the adequacy of estimated distributions, we set $\alpha = 0.05$ as a statistical significance level for the chi-square and K-S tests. Moreover,   we use $k=15$ bins for chi-square tests. As such, the critical test values for the chi-square  and  K-S tests are respectively  $CV_{\chi^2}=23.685$ and  $CV_{\text{K-S}}=0.0608$. All our numerical computations are conducted in Matlab. 

\paragraph{{\bf Experimental Results on  Distribution Estimation.}} 
Table~\ref{table:DKL-Chi2-K-SforBenchmarks}  summarizes our experimental results on  estimating the distributions of our benchmark programs. 
The first column of  Table~\ref{table:DKL-Chi2-K-SforBenchmarks} lists the name of the benchmark. 
The se\-cond column of
Table~\ref{table:DKL-Chi2-K-SforBenchmarks} specifies the random variable of the benchmark for which $|S_M|$ moments are derived, in order to estimate the distribution of the respective variable. The number $|S_M|$ of moments used for estimating distributions are given in column~3 of Table~\ref{table:DKL-Chi2-K-SforBenchmarks}.
Columns~4--5 of 
Table~\ref{table:DKL-Chi2-K-SforBenchmarks}
respectively give the resulting chi-square test values for the ME and GC expansion estimations of the pdf of the respective PP variable of the benchmark. 
In addition to the chi-square test values, columns~4--5 also indicate the adequacy  of our estimated distribution (as in lines~10--11 of Algorithm~\ref{fig:algo}): we use
\cmark{} to specify that the estimated pdf is an accurate estimate of the true distribution, and write  \xmark{} otherwise. 
The results reported in columns 6--7 of Table~\ref{table:DKL-Chi2-K-SforBenchmarks} are as in columns 4--5, yet by using the K-S test instead of the  chi-square test.

 Table~\ref{table:DKL-Chi2-K-SforBenchmarks} indicates that our approach in Algorithm~\ref{fig:algo}, based on GC expansion and ME, for distribution estimation  yields accurate estimates for pdfs of   continuous random variables, as assessed by either the chi-square test or the K-S test; the benchmarks of {\sc StutteringP}, {\sc Random Walk 1D}, {\sc Vasicek Model}, and {\sc PDP Model} fall in this category. 
 For estimating the pdf of discrete random variables, as in the {\sc Square} and {\sc Binomial} programs, the K-S test infers our method to be accurate, but the chi-square test does not. 
 The  GC expansion (see Section~\ref{ssec:reconstructing-distribution}) expresses a distribution as a series in terms of the normal distribution. When estimating pdfs of random variables whose distributions are markedly different from the normal, this GC expansion is not adequate. This occurs in the  {\sc Uniform} example, where the PP implements a uniform distribution. In contrast, the ME based  pdf estimation is accurate, as concluded by  both the chi-square and the K-S test. 
 %


\paragraph{{\bf Precision Evaluation of Higher-Order Moments of PP Variables.}}
In addition to effectively estimating distributions of a  PP variable $x$, in our experiments we were also interested to compare the higher-order moments of the estimated pdfs of $x$ against the exact moments of $x$. That is, we were interested to see how the estimated moments that we compute  from  the $|S_M|$ exact moments of $x$ differ from their respective, exact higher-order moments. 

For this evaluation setting, we use the loop iteration $n=100$ as before and apply the following setting: we compute the higher-order moments $M'_{\me}$ and $M'_{\gc}$ of the  estimated distributions $f_{\me}$ and $f_{\gc}$, and  compare 
them with the set $S_M$ of exact moments of $x$. In the sequel, we let  
$\E(x^i)_{\me}$ and  $\E(x^i)_{\gc}$
denote the $i$th estimated moment of $x$ computed from $f_{\me}$ and $f_{\gc}$,
and write ${\E(x^i)}$ for the $i$th exact moment of $x$ from $S_M$ (as in Algorithm~\ref{fig:algo}). 
For comparing moments, we compute  the \textit{absolute estimate error} for the $i$th moment of $x$ as the difference between the respective  estimated (either from $M'_{\me}$ or $M'_{\gc}$) and exact moments (from $S_M$), i.e. 
\begin{equation}\label{AbsoluteEstimateErrorDef}
\begin{array}{l}
  \text{AE}_{\me}=|{\E(x^i)}_{\me}-{\E(x^i)}|,\\
  \text{AE}_{\gc}=|{\E(x^i)}_{\gc}-{\E(x^i)}|.\\
  \end{array}
\end{equation}
In addition, we also compare the exact moments of $x$ against its respective moments obtained from sampling the PP (from $Sample_{Data}$ in Algorithm~\ref{fig:algo}). We write ${\E(x^i)}_{Sample}$ to denote the $i$th higher-order moment of $x$ obtained from $Sample_{Data}$. {As such, the    
 \textit{absolute sample error} between the sampled and exact moments of $x$ is derived by
\begin{equation}\label{AbsoluteSampleErrorDef}
  \text{AE}_{Sample}=|{\E(x^i)}_{Sample}-{\E(x^i)}|. 
\end{equation} 
The respective 
{\it relative errors} $\text{RE}_{\me}$ and $\text{RE}_{\gc}$ of the ME and GC estimation, as well as the relative error  $\text{RE}_{Sample}$ based on sampled data, are computed as 
\begin{equation}\label{eq:RE}
\begin{array}{l}
    \text{RE}_{\me}=\frac{\left|{\E(x^i)}_{\me}-{\E(x^i)}\right|}{\E(x^i)},\\ 
    \text{RE}_{\gc}=\frac{\left|{\E(x^i)}_{\gc}-{\E(x^i)}\right|}{\E(x^i)},\\ 
    \text{RE}_{Sample}=\frac{\left|{\E(x^i)}_{Sample}-{\E(x^i)}\right|}{\E(x^i)}.
\end{array}
\end{equation}
 }  
Table~\ref{table:AbsoluteandRelativeErrors} summarizes our experiments on evaluating the precision of sampled and estimated moments against exact moments, for each PP in Table~\ref{table:DKL-Chi2-K-SforBenchmarks}. 
Columns~1--3 of Table~\ref{table:AbsoluteandRelativeErrors} correspond to columns~1--3 of Table~\ref{table:DKL-Chi2-K-SforBenchmarks}. 
Column~4 lists the the order of the moment of the random variable in column~2: for each $i$th moment, we give its exact value (column~5), as well as its absolute~\eqref{AbsoluteEstimateErrorDef} and relative estimate errors~\eqref{eq:RE} (in parentheses) using $Sample_{Data}$ (column~6), ME (column~7) and GC expansion (column~8).


\begin{table}[t!]
    \centering
    \resizebox{\columnwidth}{!}{
 \begin{tabular}{c|c|c|c|l|l|l|l}
\toprule
  {\bf Program}  & Var &$|S_{\text{M}}|$ & Moment &   Exact Moment   & AE$_{Sample}$ (RE$_{Sample}$)  & AE$_{\me}$ (RE$_{\me}$) & AE$_{\gc}$ (RE$_{\gc}$)  \\
                         \midrule
 \multirow{3}{*}{StutteringP}& & &1& 210 & $5.69\times 10^{-2} (0.027\%)$  &$8.08\times 10^{-5} (0.00003\%)$&$4.71\times 10^{-5} (0.00002\%)$\\
    & $s$&2&2& $4.4405 \times 10^4 $ &  $2.39 \times 10^{1} (0.053\%)$& $1.75 \times 10^{-2} (0.00003\%)$&$1.75 \times 10^{-2} (0.00003\%)$\\
    &&&3& $9.4536\times 10^6$  & $6.78 \times 10^{3} (0.0.072\%)$ & $9.21 \times 10^{1} (0.000974\%)$ & $9.25 \times 10^{1} (0.000978\%)$\\
    &&&4& $2.0260\times 10^9$  & $1.48 \times 10^{6} (0.073\%)$& $7.38 \times 10^{4} (0.00364\%)$   & $7.42 \times 10^{4} (0.00366\%)$\\
    &&&5& $4.3705\times 10^{11}$  & $2.17 \times 10^{8} (0.050\%)$ & $3.82 \times 10^{7} (0.00873\%)$ & $3.83 \times 10^{7} (0.00876\%)$ \\
    &&&6&  $9.4884\times 10^{13}$ & $5.51 \times 10^{9} (0.0058\%)$ & $1.60 \times 10^{10} (0.0168\%)$  & $1.60 \times 10^{10} (0.0168\%)$\\
    &&&7& $2.0729\times 10^{16}$  & $2.10 \times 10^{13} (0.101\%)$& $5.88 \times 10^{12} (0.0284\%)$  & $5.90 \times 10^{12} (0.0285\%)$\\
    &&&8& $4.5570\times 10^{18}$  &$1.12 \times 10^{16} (0.245\%)$ &  $1.99 \times 10^{15} (0.0438\%)$  & $2.00 \times 10^{15} (0.0439\%)$
    \\\midrule
 \multirow{3}{*}{Square}&&&1& 10100 & $1.66\times 10^1 (0.16\%)$  &  $1.11\times 10^{0} (0.011\%)$  & $7.97\times 10^{-3} (0.00007\%)$ \\
    &$y$&2&2& $1.0602 \times 10^8$ & $2.31\times 10^5 (0.22\%)$ & $2.30\times 10^3 (0.00217\%)$   & $1.64\times 10^{2} (0.00015\%)$ \\
    &&&3& $1.1544\times 10^{12}$  &$1.27\times 10^{9} (0.11\%)$  &  $1.26\times 10^7 (0.0011\%)$  &$2.39\times 10^{9} (0.2072\%)$\\
    &&&4& $1.3012\times 10^{16}$  & $2.98\times 10^{13} (0.23\%)$& $1.99\times 10^{12} (0.0153\%)$   &$9.79\times 10^{13} (0.7581\%)$ \\
    &&&5& $1.5157\times 10^{20}$  & $1.32\times 10^{18} (0.87\%)$& $8.83\times 10^{16} (0.0583\%)$    &$2.61\times 10^{18} (1.7505\%)$
    \\\midrule
\multirow{3}{*}{Binomial}&&&1& 50 & $2.95\times 10^{-1} (0.59\%)$  &  $5.19\times 10^{-5} (0.000051\%)$ & $3.16\times 10^{-3} (0.0063\%)$\\
    &$x$&2&2&   2525  & $2.82\times 10^{1} (1.12\%)$ & $2.74\times 10^{-3} (0.000108\%)$  & $1.87\times 10^{-1} (0.0074\%)$ \\
    &&&3& 128750  & $2.04\times 10^{3} (1.59\%)$ & $1.43\times 10^{-1} (0.000111\%)$  & $1.22\times 10^{1} (0.0095\%)$ \\
    &&&4&  $6.6268\times 10^{6}$  &$1.32\times 10^{5} (2.00\%)$ & $3.66\times 10^{0} (0.000055\%)$  & $8.22\times 10^{2} (0.0125\%)$ \\
    &&&5&  $3.4421\times 10^{8}$ &$8.08\times 10^{6} (2.35\%)$ &  $5.52\times 10^{2} (0.000160\%)$ & $5.53\times 10^{4} (0.0161\%)$ \\
    &&&6& $1.8038\times 10^{10}$ & $4.67\times 10^{8} (2.64\%)$& $1.20\times 10^{5} (0.000664\%)$   & $3.66\times 10^{6} (0.0203\%)$\\
    &&&7&  $9.5354\times 10^{11}$ &$2.74\times 10^{10} (2.87\%)$ &  $1.51\times 10^{7} (0.001587\%)$  & $2.38\times 10^{8} (0.0250\%)$\\
    &&&8&  $5.0830\times 10^{13}$ &$1.54\times 10^{12} (3.04\%)$ &  $1.54\times 10^{9} (0.0030363\%)$   & $1.51\times 10^{10} (0.0298\%)$
         \\\midrule
 \multirow{3}{*}{RandomWalk1D}&&&1& 20 & $1.57\times 10^{-1} (0.79\%)$  &$1.83\times 10^{-6} (0.000009\%)$ & $4.44\times 10^{-3} (0.022\%)$\\
    &$x$&2&2& $4.2933\times 10^{2} $ &  $6.78 \times 10^{0} (1.58\%)$& $6.76 \times 10^{-5} (0.00001\%)$ & $1.90\times 10^{-1} (0.0044\%)$\\
    &&&3& $9.7516\times 10^{3}$  & $2.45 \times 10^{2} (2.51\%) $& $8.40 \times 10^{0} (0.09\%)$  & $5.61\times 10^{-1} (0.0057\%)$\\
    &&&4&  $2.3230\times 10^{5}$ & $8.38\times 10^{3} (3.61\%)$&  $6.48\times 10^{2} (0.28\%)$  & $3.54\times 10^{2} (0.15\%)$\\
    &&&5& $5.7681\times 10^{6}$  & $2.79\times 10^{5} (4.48\%)$&  $3.37\times 10^{4} (0.58\%)$ &$2.32\times 10^{4} (0.40\%)$ \\
    &&&6&  $1.4858\times 10^{8}$ & $9.19\times 10^{6} (6.19\%)$&  $1.48\times 10^{6} (0.98\%)$ & $1.12\times 10^{6} (0.75\%)$ \\
    &&&7& $3.9565\times 10^{9}$  & $3.03\times 10^{8} (7.66\%)$& $5.94\times 10^{7} (1.48\%)$  & $4.72\times 10^{7} (1.18\%)$ \\
    &&&8& $1.0857\times 10^{11}$  & $1.01\times 10^{10} (9.26\%)$&  $2.26\times 10^{9} (2.04\%)$  & $1.85\times 10^{9} (1.67\%)$
    \\\midrule
 \multirow{3}{*}{Uniform(0,1)}&&&1& 0.5 & $5.92\times 10^{-3} (1.18\%)$  & $1.47\times 10^{-9} (3\times 10^{-7}\%)$ & $3.28\times 10^{-2} (7.03\%)$\\
    &$u$&6&2& 0.333333 & $4.94\times 10^{-3} (1.48\%)$ &   $1.66\times 10^{-5} (0.00499\%)$& $3.40\times 10^{-2} (11.36\%)$\\
    &&&3& 0.25  & $4.57\times 10^{-3} (1.83\%)$ & $2.50\times 10^{-5} (0.00998\%)$  & $3.46\times 10^{-2} (16.06\%)$ \\
    &&&4&  0.20  &$4.57\times 10^{-3} (2.28\%)$ &  $3.33\times 10^{-5} (0.0166\%)$  & $3.43\times 10^{-2} (20.69\%)$\\
    &&&5&  0.166667  & $4.66\times 10^{-3} (2.80\%)$&   $4.18\times 10^{-5} (0.0249\%)$ &$3.35\times 10^{-2} (25.18\%)$ \\
    &&&6&  0.142857 &$4.74\times 10^{-3} (3.32\%)$ &   $4.99\times 10^{-5} (0.0349\%)$ & $3.25\times 10^{-2} (29.51\%)$\\
    &&&7&  0.125 &$4.76\times 10^{-3} (3.81\%)$ &     $5.82\times 10^{-5} (0.00465\%)$ &$3.15\times 10^{-2} (33.65\%)$\\
    &&&8&  0.1111111 &$4.73\times 10^{-3} (4.25\%)$ &  $6.66\times 10^{-5} (0.00599\%)$ &  $3.04\times 10^{-2} (37.60\%)$\\\midrule
\multirow{3}{*}{Vasicek Model}&&&1& 0.20 & $2.16\times 10^{-3} (1.08\%)$  &$6.63\times 10^{-11} (3.32\times 10^{-8}\%)$ &$1.40 \times 10^{-8} (7.00\times 10^{-6}\%)$\\
     &$r$&2& 2& $0.0933$ &  $5.91 \times 10^{-3} (6.33\%)$& $9.67\times 10^{-11} (1.04\times 10^{-7}\%)$ &$2.15 \times 10^{-8} (2.31\times 10^{-5}\%)$ \\
     
    &&&3& $0.0400$  & $3.13 \times 10^{-3} (7.83\%)$& $2.00 \times 10^{-8} (5.00\times 10^{-5}\%)$& $3.32 \times 10^{-8} (8.30\times 10^{-5}\%)$\\
    
    &&&4& 0.0229 & $ 2.54\times 10^{-3} (11.09\%) $ &   $4.02 \times 10^{-8} (1.75\times 10^{-4}\%)$ & $5.11\times 10^{-8} (2.23\times 10^{-4}\%)$ \\
    &&&5& 0.0131  & $ 1.83\times 10^{-3} (13.93\%) $ &   $7.11 \times 10^{-8} (5.42\times 10^{-4}\%)$ & $7.89\times 10^{-8} (6.01\times 10^{-4}\%)$\\
    &&&6& 0.0087& $ 1.61\times 10^{-3} (18.46\%) $&  $1.15 \times 10^{-7} (1.32\times 10^{-3}\%)$ & $1.22\times 10^{-7} (1.39\times 10^{-3}\%)$\\
    &&&7& 0.0059  &$ 1.40\times 10^{-3} (23.49\%) $ &   $1.83 \times 10^{-7} (3.07\times 10^{-3}\%)$ & $1.88\times 10^{-7} (3.16\times 10^{-3}\%)$\\
    &&&8&  0.0044 & $ 1.33\times 10^{-3} (29.86\%) $&   $2.86 \times 10^{-7} (6.43\times 10^{-3}\%)$ & $2.91\times 10^{-7} (6.53\times 10^{-3}\%)$
    \\\midrule
 \multirow{3}{*}{PDP Model}&&&1& $1.1885\times 10^{3}$ & $1.60\times 10^{1}  (1.35\%)$  &$2.84\times 10^{-1}  (0.024\%)$ &$5.74\times 10^{0}  (0.48\%)$ \\
    &$x$&3&2& $1.4767\times 10^{6}$ &  $3.93\times 10^{4} (2.66\%)$& $3.95\times 10^{2} (0.027\%)$ &$1.16\times 10^{4} (0.78\%)$ \\
    &&&3& $1.8981\times 10^{9}$  & $7.37 \times 10^{7} (3.88\%)$& $3.19\times 10^{6}(0.168\%)$ & $2.35\times 10^{7} (1.23\%)$\\
     &&&4& $2.5058\times 10^{12}$ & $1.28\times 10^{11}(5.01\%)$ &  $1.64\times 10^{10}$ (0.650\%)  & $4.85\times 10^{10} (1.90\%)$\\
     &&&5& $3.3804\times 10^{15}$  &$2.04\times 10^{14}(6.03\%)$  & $ 4.97\times 10^{13} (1.450\%)$ & $1.00\times 10^{14} (2.87\%)$\\
        \bottomrule
\end{tabular}
}
\vspace{2mm} 
 \caption{Precision evaluation of higher-order moments  using $|S_M|$ exact moments.  
 }
\label{table:AbsoluteandRelativeErrors}
\end{table}




{ 
Table~\ref{table:AbsoluteandRelativeErrors} gives practical  evidence of the accuracy of estimating the pdf, and hence moments, using Algorithm~\ref{fig:algo}. The absolute and relative errors listed in Table~\ref{table:AbsoluteandRelativeErrors}  show  that
we gain higher precision when computing moments from the estimated pdfs using ME and/or GC expansion when compared to the moments  calculated
using sampled data.  
The accuracy of moments calculated from sampled data depends on the quality of the sampling process, which in turn  depends on the number of samples ($e$) and number of loop iterations ($n$). Our results in Table~\ref{table:AbsoluteandRelativeErrors} indicate that computing  moments from estimated pdfs provides a more  accurate and time-efficient alternative to estimating moments by sampling. 
%
}

\section{Conclusion}\label{section:Conclusion}
We present an algorithmic approach to  estimate the distribution of the random variables of a PP using a finite number of its moments.   Our estimates are based on Maximum Entropy and Gram-Charlier expansion. The accuracy of our estimation is assessed with the  chi-square and Kolmogorov-Smirnov  statistical tests. Our evaluation combines static analysis methods for the  computation of exact moments of a PP random variable $x$ with the aforementioned statistical techniques to produce estimates of the distribution of $x$. 
Extending our approach 
to support probabilistic inference  and quantify the loss of precision in the estimation are  future research directions. 
\bibliographystyle{splncs04}
\bibliography{refs1}




\end{document}